\documentclass[prl,aps,preprint,showpacs]{revtex4}

\usepackage{graphicx}
\usepackage{subfigure}
\usepackage{amsfonts}
\usepackage{amssymb}
\usepackage{amsmath}
\usepackage{hyperref}
\begin{document}

\title{Terahertz Response of Acoustically-Driven Optical Phonons}
\date{\today}

\author{R.~H. Poolman}
\author{E.~A. Muljarov}
\author{A.~L. Ivanov}
\affiliation{Department of Physics and Astronomy, Cardiff
University, Cardiff CF24 3AA, United Kingdom.}

\begin{abstract}

The manipulation of TO-phonon polaritons and the terahertz (THz)
light field associated with them by means of an ultra-sound
acoustic wave is proposed and illustrated by calculating the
TO-phonon-mediated THz response of acoustically-pumped CuCl and
TlCl crystals. We show the high-contrast acoustically-induced
change of the THz reflectivity alongside with multiple THz Bragg
replicas, which are associated with the infrared-active TO-phonon
resonance driven by the ultrasonic wave. The effect, which stems
from phonon anharmonicity, refers to an operating acoustic
intensity $I_{\rm ac} \sim 1-100$\,kW/cm$^2$ and frequency
$\nu_{\rm ac} \sim 0.1 - 1$\,GHz, with possible applications in
THz spectroscopy.

\end{abstract}


\maketitle

Since the pioneering work by Kun Huang~\cite{Huang1951}, the
physics of infrared polaritons associated with transverse optical
(TO) phonons has emerged as a well-established discipline.
Recently, room-temperature polaritonics was implemented for
processing and coherent control of the THz light
field~\cite{Nelson2002,Nelson2003}. In addition to conventional
infrared spectroscopy and continuous-wave Raman scattering
experiments, broadband THz time-domain spectroscopy has been
developed and successfully applied to characterize picosecond (ps)
and sub-ps dynamics of infrared-active TO-phonons~\cite{Han2001}
and to visualize the polariton dispersion associated with these
vibrational modes~\cite{Kojima2003}. Furthermore, the THz
polariton spectra allow to study unusual lattice dynamics, e.g.,
in ferroelectrics \cite{Bakker1998} and negative thermal expansion
compounds \cite{Hancock2004,Chaplot2005}. However, the
spectrally-resolved control of the THz electromagnetic field
associated with infrared polaritons still remains a very
challenging task of far-infrared spectroscopy.

In this Letter we propose an acoustic modulation of TO-phonon
polaritons to drastically change their optical response in the THz
band. Phonon anharmonicity, which can be large in some dielectric
and semiconductor materials and particularly strong for soft
TO-phonon modes, leads to the coupling between a coherent
(pumping) acoustic wave (AW) and infrared-active TO-phonons. In
this case one deals with an {\it acoustically-induced
Autler-Townes effect}, which gives rise to spectral gaps
$\Delta_N$ in the THz polariton spectrum of AW-driven TO-phonons,
and is akin to the acoustical Stark effect for excitons
\cite{Ivanov2001}. The AW-induced gaps $\Delta_N$, which open up
in the polariton spectrum and develop with increasing acoustic
intensity $I_{\rm ac}$, are due to the $N$th-order resonant
acoustic phonon transitions within the polariton dispersion
branches. These forbidden-energy gaps strongly modify the optical
response of TO-phonon polaritons and make possible the effective
AW manipulation of the THz field.

The spectrally-resolved AW control of the THz field propagation
can also be interpreted in terms of Bragg diffraction of infrared
polaritons by the pumping AW: For the first time we analyze the
use of phonon anaharmonicity to create an acoustically-induced
Bragg grating. In this case the contrast of the AW grating is
dictated by the efficiency of the scattering channel ``TO-phonon
$\pm$ acoustic phonon (two acoustic phonons) $\leftrightarrow$
TO-phonon'' for cubic (quartic) phonon anharmonicity. Thus the
scattering of THz light is mediated and strongly enhanced by the
TO-phonon resonance. This results in an anomalously short
interaction length needed for formation of the Bragg replicas.
Possible applications of governing infrared polaritons by using an
ultrasonic acoustic field include frequency-tunable THz detectors
and filters, Bragg switchers and frequency converters.

The Hamiltonian of infrared polaritons coherently driven by a cw
bulk acoustic wave $\{ {\bf k},\omega_{\rm ac}(k) \}$ is given by
\begin{eqnarray}
H = H_0 + \sum_{\bf p} \Big[ 2 \tilde{m}_4 I_{\rm ac} b_{\bf
p}^{\dag} b_{\bf p} &+& \big( m_3 I_{\rm ac}^{1/2} e^{-i
\omega_{\rm
ac} t} b_{\bf p}^{\dag} b_{\bf p-k} + {\rm H.\,c.} \big)\nonumber \\
&+& \big( m_4 I_{\rm ac} e^{- 2i \omega_{\rm ac} t} b_{\bf
p}^{\dag} b_{{\bf p}-2{\bf k}} + {\rm H.\,c.} \big) \Big] \, ,
\label{ham}
\end{eqnarray}
with $H_0$ the conventional polariton Hamiltonian of
infrared-active TO-phonons~\cite{RomeroRochin1999}, $b_{\bf p}$
the TO-phonon operator, $\omega_{\rm ac} = 2\pi \nu_{\rm ac} =
v_{\rm s} k$, $v_{\rm s}$ the sound velocity, and $m_{3}$ ($m_{4}$
and $\tilde{m}_4$) the matrix element associated with cubic
(quartic) anharmonicity. The macroscopic equations, which describe
the control of THz polaritons by applying the acoustic field of an
arbitrary profile, $I_{\rm ac} = I_{\rm ac}({\bf r},t)$, are
\begin{eqnarray}
&&\bigg[ \frac{\varepsilon_{\rm b}}{c^2}
\frac{\partial^2}{\partial t^2} - \nabla^2 \bigg] {\bf E}({\bf
r},t) = - \frac{4 \pi}{c^2}\frac{\partial^2}{\partial t^2} \, {\bf
P}({\bf r},t) \, , \label{macroE} \\
&&\bigg[\frac{\partial^2}{\partial t^2} + 2 \gamma_{\rm TO}
\frac{\partial}{\partial t} + \Omega^2_{\rm TO} + 4 \Omega_{\rm
TO} \tilde{m}_4 I_{\rm ac}({\bf r},t) + 4 \Omega_{\rm TO} m_3
I^{1/2}_{\rm ac}({\bf r},t) \cos(\omega_{\rm ac}t
- {\bf kr}) \nonumber \\
&&\ \ \ \ \ \ \ + \ 4 \Omega_{\rm TO} m_4 I_{\rm ac}({\bf r},t)
\cos(2\omega_{\rm ac}t - 2{\bf kr}) \bigg] {\bf P}({\bf r},t) =
\frac{\varepsilon_{\rm b}}{4 \pi} \Omega_{\rm R}^2 \, {\bf E}({\bf
r},t) \, , \label{macroP}
\end{eqnarray}
where ${\bf E}$ and ${\bf P}$ stand for the light field and
TO-phonon polarization, respectively, $\Omega_{\rm TO}$ is the
TO-phonon frequency, $\Omega_{\rm R}$ is the polariton Rabi
frequency, $\varepsilon_{\rm b}$ is the background dielectric
constant in the infrared, and $\gamma_{\rm TO}$ is the damping
rate of TO-phonons, mainly due to their decay into
short-wavelength acoustic phonons.
Equations~(\ref{macroE})-(\ref{macroP}) refer to simple cubic
lattices with spatially isotropic long-wavelength anharmonicity
and optical response. For cw acoustic excitations, when $I_{\rm
ac}({\bf r},t) = I_{\rm ac} = $\,const.,
Eqs.\,(\ref{macroE})-(\ref{macroP}) yield the same quasi-energy
spectrum as that of the quadratic Hamiltonian (\ref{ham}). If
$I_{\rm ac} = 0$, Eqs.\,(\ref{macroE})-(\ref{macroP}) reduce to
the standard TO-phonon polariton equations~\cite{Huang1951}. The
forth term on the left-hand side (l.h.s.) of the polarization
Eq.\,(\ref{macroP}) yields a Stark shift of the TO-phonon
frequency, which is $\propto I_{\rm ac}$ and is associated with
the quartic phonon nonlinearity. The last two terms on the l.h.s.
of Eq.\,(\ref{macroP}) give rise to the Bragg spectrum of
TO-phonon polaritons driven by the AW.

\begin{figure}[t]
\includegraphics[angle=270,width=0.65\linewidth]{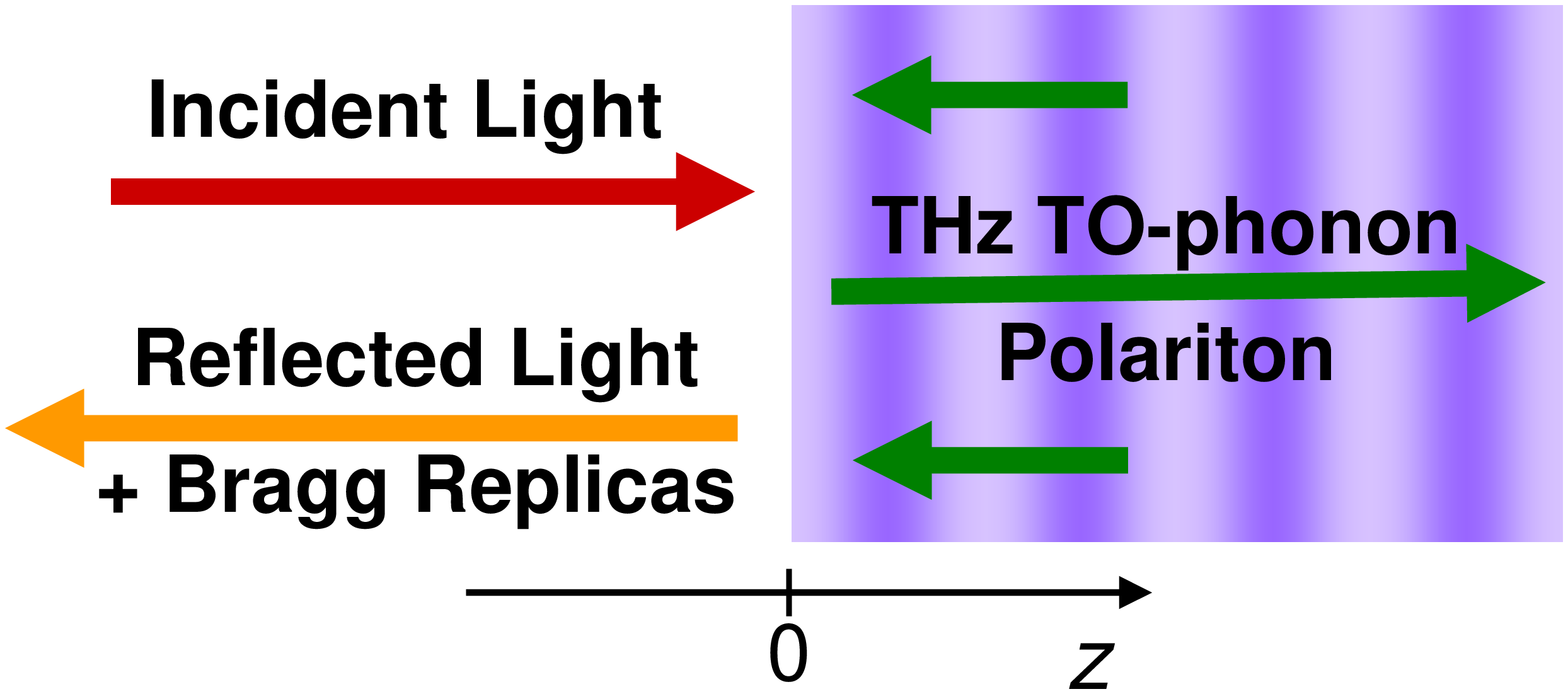}
\caption{(Color online) Schematic of optical excitation and
backward Bragg scattering of an acoustically-driven THz polariton.
Vertical stripes symbolize the propagating bulk acoustic wave.}
\end{figure}

The matrix elements associated with cubic and quartic
anharmonicity are $m_3 = 6 [v_0/(\hbar^3 v_{\rm s}^2
k)]^{1/2}V_3({\bf k},{\bf p}-{\bf k},-{\bf p})$, $m_4 = 12
[v_0/(\hbar^2 v_{\rm s}^2 k)][ V_4({\bf k},{\bf k},{\bf p}-2{\bf
k},-{\bf p}) + V_4({\bf k},{\bf k},-{\bf p},{\bf p}-2{\bf k})]$,
and $\tilde{m}_4 = 24 [v_0/(\hbar^2 v_{\rm s}^2 k)] V_4({\bf
k},-{\bf k},{\bf p},-{\bf p})$, with $v_0$ a volume of the
primitive cell. The potential $V_3$ is given by $V_3({\bf k},{\bf
p}-{\bf k},-{\bf p}) = (1/6) [\hbar^3 /(8 \omega_{\rm ac}
\Omega_{\rm TO}^2)]^{1/2} \Phi^{(3)}({\bf k},{\bf p}-{\bf k},-{\bf
p})$, where $\Phi^{(3)}$ is proportional to the Fourier transform
of the third-order derivative of the interatomic potential. There
is a similar expression for the forth-order potential, $V_4
\propto \Phi^{(4)}$ (the explicit formulae for $V_{3,4}$ and
$\Phi^{(3,4)}$ are given, e.g., in
Refs.\,\cite{Maradudin1962,Cowley1968}). In the long-wavelength
limit $k,p \ll 1/a_0$ ($a_0$ is the lattice constant), relevant to
the optics of TO-phonon polaritons, $\Phi^{(3,4)}$ are
well-approximated by $\Phi^{(3)} = (ka_0) C_3$ and $\Phi^{(4)} =
(ka_0)^2 C_4$~\cite{Srivastava1990}. The anharmonicity constants
$C_{3,4}$ can either be calculated by using modern {\it ab initio}
methods~\cite{Debernardi1995,Deinzer2004} or evaluated from
experimental data available for some anharmonic
crystals~\cite{Cardona1999,Cardona2001,Lowndes1971}. The above
approximation of $\Phi^{(3,4)}$ also leads to the $k$-independent
matrix elements $m_3$ and $\tilde{m}_4 = m_4$ in Eqs.\,(\ref{ham})
and (\ref{macroP}).

Recently, the third-order coupling constant $V_3$ was inferred for
some zincblende-type semiconductors (GaP, CuCl, CuBr, and
$\beta$-ZnS) which exhibit strong and dominant cubic
anharmonicity~\cite{Cardona1999,Cardona2001}. The used
experimental data refers to the decay of a long-wavelength
TO-phonon into short-wavelength longitudinal (LA) and transverse
(TA) acoustic phonons. To adapt the inferred values of $V_3$ to
$m_3$ in Eq.\,(\ref{ham}) and Eqs.\,(\ref{macroE})-(\ref{macroP}),
we use the Leibfried-Ludwig approximation~\cite{Leibfried1961}.
The parameter $m_3 I_{\rm ac}^{1/2}$, which controls
cubic-anharmonicity-mediated manipulation of THz polaritons by
means of a bulk TA wave, is evaluated for CuCl as $m_3 I_{\rm
ac}^{1/2} \simeq 0.26$\,meV (63\,GHz) for $I_{\rm ac} =
1$\,kW/cm$^2$. In thallium halides (TlCl and TlBr), quartic
anharmonicity is dominant with positive values of $V_4$. With the
Leibfried-Ludwig approximation, one can evaluate the control
parameter $m_4 I_{\rm ac}$ from the available experimental data on
the real part of the TO-phonon self-energy~\cite{Lowndes1971}. For
TlCl driven by a bulk TA wave we get $m_4 I_{\rm ac} \simeq
0.4$\,meV (97\,GHz) for $I_{\rm ac} = 100$\,kW/cm$^2$. Note that
$m_3$ and $m_4$ scale to the sound velocity as $v_{\rm s}^{-3/2}$
and $v_{\rm s}^{-3}$, respectively, so that the use of surface AWs
considerably reduces the operating acoustic intensity.

We examine the optical response of an acoustically-driven
TO-phonon polariton in a one-dimensional geometry, when a
semiconductor occupies the half space $z>0$, and a normally
incident light field of frequency $\omega$ induces a THz polariton
propagating collinearly to the pumping AW (see Fig.\,1). In this
case, apart from reflectivity at the same frequency $\omega$,
down-converted Bragg replicas at $\omega + n \omega_{\rm ac}$ ($n
= -1, -2, ...$) arise in the reflection spectrum, due to
acoustically-induced backward scattering of the polariton. In
order to calculate the multiple Bragg replicas, we develop an
approach more advanced than that used in conventional
acousto-optics. For the latter, the acousto-optical
susceptibilities are so weak that usually only one Bragg replica
$n = +1$ or $-1$ is seen. In contrast, the TO-phonon resonance
mediates and considerably enhances the coupling between the
optical and acoustic fields, so that generally one has to take
into account the whole series of the Bragg replicas and
multi-phonon transitions, thus treating the problem
non-perturbatively.

\begin{figure}[t]
\includegraphics[angle=0,width=0.5\linewidth]{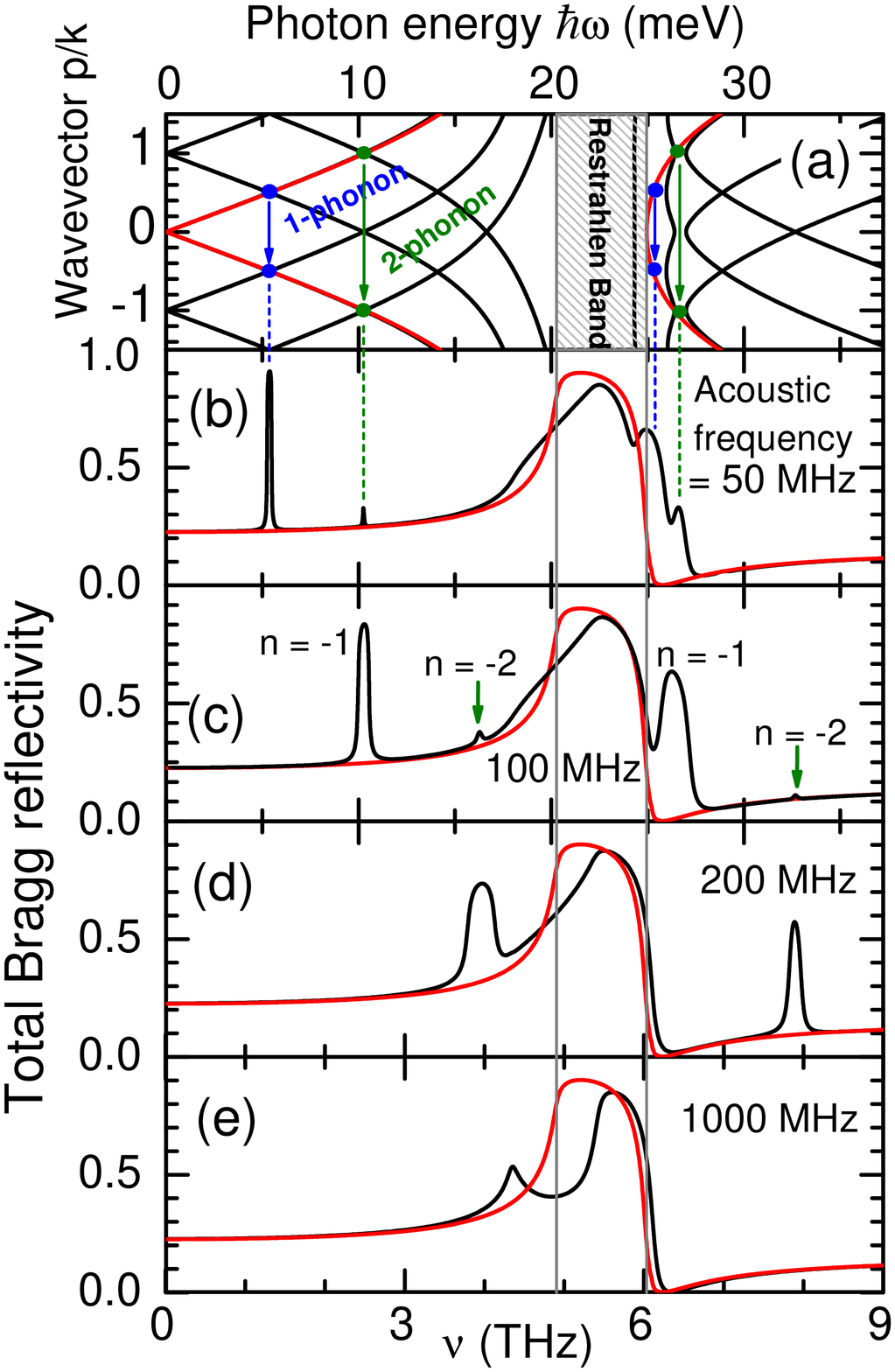}
\caption{(Color online) (a) The dispersion $\omega = \omega(p)$ of
THz polaritons in CuCl driven by the bulk TA-wave of $\nu_{\rm
ac}=50$\,MHz (black lines) and the bare polariton spectrum (red
lines). The wavevector $p$ is normalized to the acoustic
wavevector $k$. The arrows highlight the $N$-TA-phonon resonant
coupling between the polariton states ($N=1,2$). (b)-(e) The
calculated total Bragg reflectivity (black lines), $R = \sum_n
|r_n|^2$, against the light frequency $\nu = \omega/(2 \pi)$ for
$\nu_{\rm ac} = 50$\,MHz, 100\,MHz, 200\,MHz, and 1\,GHz,
respectively. The bare polariton reflectivity is shown by the red
lines. The Bragg signals labelled in (c) as $n=-1$ and $n=-2$ are
mainly due to $|r_{-1}|^2$ and $|r_{-2}|^2$, respectively. $I_{\rm
ac}= 25$\,kW/cm$^{2}$, $\hbar \Omega_{\rm TO} = 20.28$\,meV,
$\hbar \Omega_{\rm R} = 14.53$\,meV, and $\hbar \gamma_{\rm TO} =
0.2$\,meV.}
\end{figure}

The quasi-energy polariton spectrum $\omega  = \omega(p)$,
calculated for real-valued $p$ (quasiparticle solution) by solving
Eqs.\,(\ref{macroE})-(\ref{macroP}) with $m_4 = 0$, is plotted in
Fig.\,2\,(a) for CuCl parametrically driven by the TA wave of
frequency $\nu_{\rm ac} = 50$\,MHz and constant intensity $I_{\rm
ac} = 25$\,kW/cm$^2$. The spectrum, which arises from spatial and
temporal modulation of the crystal lattice, can be interpreted in
terms of the Brillouin zone picture, with acoustically-induced
energy gaps $\Delta_N \propto I_{\rm ac}^{N/2}$ due to the
$N$-phonon resonant transitions within the polariton dispersion
branches. The spectral positions of the transitions are indicated
in Fig.\,2\,(a) by the vertical arrows. For the frequency scale
used in Fig.\,2\,(a), only the gaps $\Delta_{N=1}$ and
$\Delta_{N=2}$ in the upper polariton branch are clearly seen.

From the air side, $z < 0$ (see Fig.\,1), the light field is given
by
\begin{equation}
E(z < 0,t) = e^{iq_0z} e^{-i\omega t} + \sum_{n}{r_n e^{-iq_nz}
e^{-i(\omega + n \omega_{\rm ac})t}} \, , \label{left}
\end{equation}
where $q_{n}=(\omega + n \omega_{\rm ac})/c$ with $-n_{\rm max}
\leqslant n \leqslant n_{\rm max}$ (we proceed up to $n_{\rm
max}=60$) and $r_n$ stands for the amplitude of the outgoing Bragg
replica $n$ normalized to the unity amplitude of the incoming
light wave. The electric field propagating in the crystal ($z >
0$) is
\begin{equation}
E(z > 0,t) = \sum_{n,j}{A_j E_{nj} e^{i(p_j + n k)z - i(\omega + n
\omega_{\rm ac})t}} \, . \label{right}
\end{equation}
Here, $p_j = p_j(\omega)$ is the wavevector associated with the
quasi-energy dispersion branch $j$ ($n_{\rm max} \leqslant j
\leqslant n_{\rm max}$) of the acoustically-driven polariton,
$E_{nj}$ are the corresponding normalized eigenvectors, and $A_j$
are the eigenmode amplitudes. Both $p_j = p_j(\omega)$ and
$E_{nj}$ are the forced-harmonic solutions of
Eqs.\,(\ref{macroE})-(\ref{macroP}) for real-valued frequency
$\omega$. The exponential on the r.h.s. of Eq.\,(\ref{right}) as
well as the basic relationships $p_{j + s}(\omega) = p_j(\omega -
s \omega_{\rm ac}) + sk$ and $E_{n,j+s}(\omega)=E_{n+s,j}(\omega -
s \omega_{\rm ac})$, with integer $s$, reflect the acoustic
wavevector and frequency translational invariance of the
quasi-energy spectrum. The Maxwellian boundary conditions at $z=0$
together with Eqs.\,(\ref{left})-(\ref{right}) yield:
\begin{equation}
\delta_{n,0} + r_n = \sum_j A_j E_{nj} \,, \ \ \ \ \ \
\delta_{n,0} - r_n = \sum_j A_j E_{nj} \frac{p_j + nk}{q_n} \, .
\label{mbc}
\end{equation}
The set of $2(2n_{\rm max} + 1)$ linear Eqs.\,(\ref{mbc})
determines $r_n$ and $A_j$.

\begin{figure}[t]
\includegraphics[angle=270,width=0.5\linewidth]{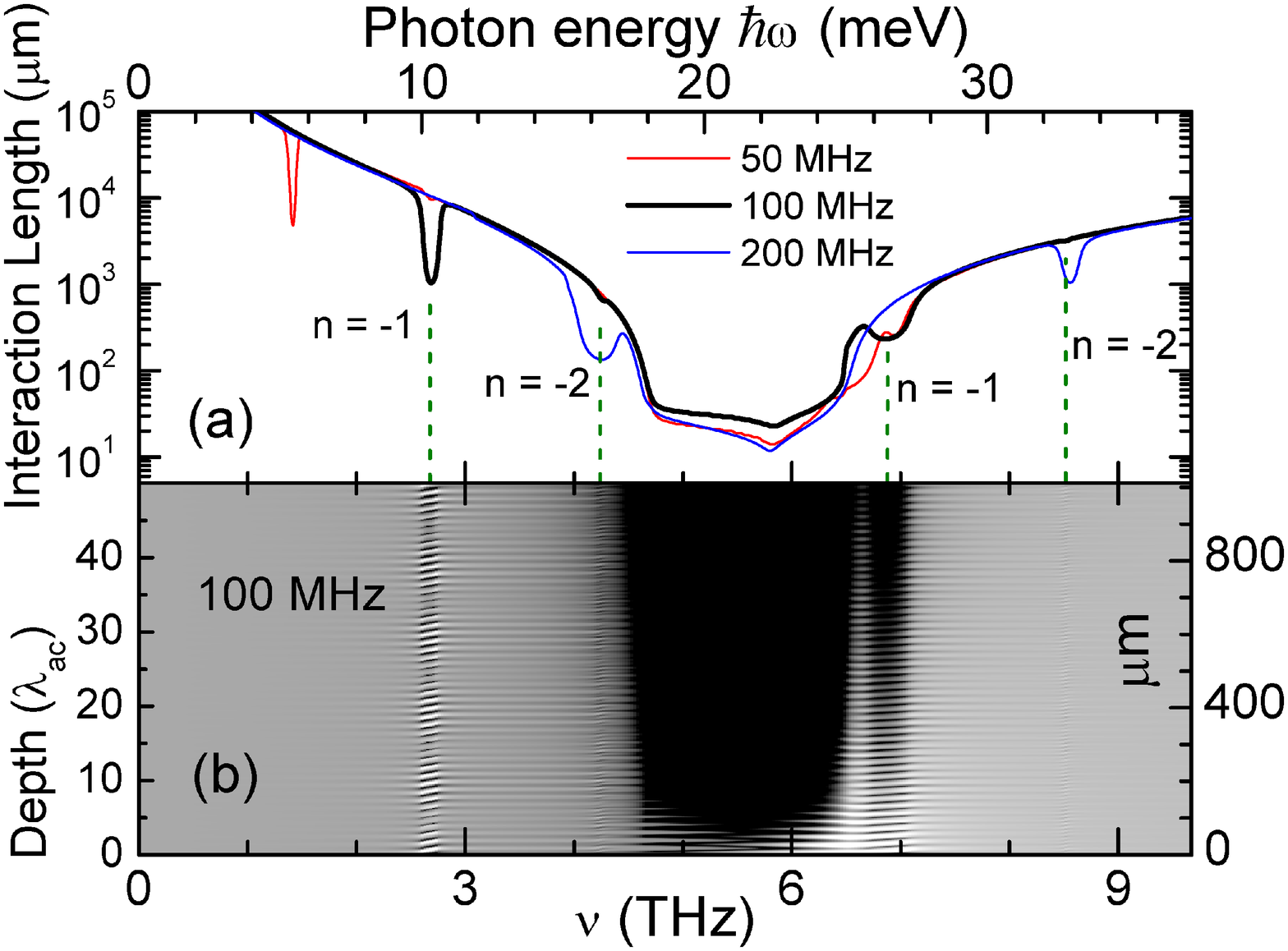}
\caption{(Color online) (a) The interaction length between the THz
polariton and TA pumping wave in CuCl, $\ell_{\rm int} = \ell_{\rm
int}(\omega)$, calculated for $I_{\rm ac} = 25$\,kW/cm$^{2}$ and
$\nu_{\rm ac} = 50$\,MHz, 100\,MHz, and 200\,MHz. (b) The electric
field profile $|E(z>0,\omega)|^{2}$ evaluated for $\nu_{\rm ac} =
100$\,MHz ($\lambda_{\rm ac} = 20.2\,\mu$m). The color scale is
logarithmic with black color corresponding to $E \rightarrow 0$.}
\end{figure}

In Figs.\,2\,(b)-(e) we compare the calculated total Bragg
reflectivity, $R = R(\omega) = \sum_n |r_n|^2$ (black solid
lines), with the reflectivity of the acoustically unperturbed THz
polariton, $R^{(0)} = R^{(0)}(\omega)$ (red solid lines). The
sharp spikes in the Bragg spectrum of the acoustically-driven CuCl
crystal are clearly seen for the one-TA-phonon and two-TA-phonon
transitions, both for the upper (UP) and lower (LP) polariton
branches [see Fig.\,2\,(a) against Fig.\,2\,(b)]. For $\nu_{\rm
ac} \sim 30 - 300$\,MHz, the backward scattered Bragg replica
$|r_{-1}|^2$ peaks at the energy of the one-phonon transition and
is highly efficient, with $|r_{-1}|^2/R \sim 50\,\% - 70\,\%$. The
peak position and its strength are effectively tunable by changing
the frequency and intensity of the AW [see Figs.\,2\,(b)-(d)].
This can be used for the frequency down-conversion by $\omega_{\rm
ac} = 2 \pi \nu_{\rm ac}$ of the optically-induced THz polariton
and of the incident light field. Generally, the backward Bragg
scattering signals $|r_{n}|^2 \propto I_{\rm ac}^{|n|}$ ($n < 0$)
peak at the spectral position of the gaps $\Delta_{N \leqslant
|n|}$, i.e., for the light frequency $\omega = \omega_N$ which
satisfies the resonant Bragg condition $p_{j=0}(\omega_N) \simeq N
k/2$ with $N= 1,2,..,|n|$.

The Bragg signal $n=0$ appears as the AW-induced change of the
reflectivity at incident frequency $\omega$, $|r_0(\omega)|^2 -
R^{(0)}(\omega)$. The strength of the $n=0$ signal sharply
increases with decreasing detuning $|\omega - \Omega_{\rm TO}|$
from the TO-phonon resonance, so that $|r_0|^2 - R^{(0)}$ becomes
dominant over $|r_{-1}|^2$. For $\nu_{\rm ac} \gtrsim 1$\,GHz,
when the one-acoustic-phonon transition within the LP branches
occurs very close to $\Omega_{\rm TO}$, the AW-induced change of
the reflectivity [see Fig.\,2\,(e)] is completely determined by
the $n=0$ replica and has no $\nu_{\rm ac}$-down-converted
frequency components. Thus this operating mode can be used for
TO-phonon polariton deflectors and acoustically-controlled THz
filters.

The interaction length $\ell_{\rm int} = \ell_{\rm int}(\omega)$
required for the formation of the Bragg signals and thus for AW
control of the THz light field is plotted in Fig.\,3\,(a) for
various $\nu_{\rm ac}$. The sharp troughs at $\omega =
\omega_{N=1}$ and $\omega_{N=2}$ in the $\ell_{\rm int} =
\ell_{\rm int}(\omega)$ profile are due to the $n = -1$ and $-2$
Bragg replicas [see Fig.\,3\,(a)]. Figure~3\,(b) shows the light
field distribution associated with the optically-induced THz
polariton in the acoustically-driven CuCl crystal. Apart from the
broad black band [see Fig.\,3\,(b)], which corresponds to the
Restrahlen band with rather weak penetration of the light field
into the crystal, the narrow stripes of alternating color
illustrate the formation of the Bragg replicas. The interaction
length for the $n=-1$ replica at its resonant frequency $\omega =
\omega_{N=1}$ is given by
\begin{equation}
\ell_{\rm int} = \frac{\hbar k c^2}{4 m_3 I_{\rm ac}^{1/2}
\varepsilon_{\rm b}} \, \frac{(\omega_{N = 1}^2 - \Omega_{\rm
TO}^2)^2}{\omega_{N=1}^2 \Omega_{\rm TO} \Omega_{\rm R}^{2}} \, .
\label{length}
\end{equation}
Equation~(\ref{length}), which is valid for $|\Omega_{\rm TO} -
\omega_{N=1}| \gg \gamma_{\rm TO}$ and $\ell_{\rm int} k \gg 1$,
shows the resonant decrease of $\ell_{\rm int} \propto
1/\sqrt{I_{\rm ac}}$ with decreasing frequency detuning from the
TO-phonon resonance. In this case the interaction between the
light field and pumping AW is mediated by the TO-phonon resonance,
giving rise to $\ell_{\rm int}$ of only a few tens of acoustic
wavelength $\lambda_{\rm ac}$ (see Fig.\,3). This is in sharp
contrasts with conventional acousto-optics where $\ell_{\rm int}
\sim 10^3-10^4\,\lambda_{\rm ac}$ for the same operating $I_{\rm
ac}$.

Within the used nonperturbative approach, each Bragg replica $n$
integrates all $n + s - s$ TA-phonon transitions with $s \leqslant
n_{\rm max}$: With increasing $I_{\rm ac}$ the bare $n$-phonon
transitions become dressed by higher-order processes when $n+s$
phonons are emitted and $s$ phonons absorbed. For the $n=-1$
replica shown in Fig.\,2\,(b), e.g., the multiphonon transitions
$-1+1-1$, $-1+2-2$, etc. account for about 90\,\% of $|r_{-1}|^2$.
For $\nu_{\rm ac} \gtrsim 1$\,GHz, the dominant contribution to
$|r_0|^2 - R^{(0)}$ stems from $-s + s$ multi-TA-phonon
transitions with $s > 1$.

The calculated room temperature reflectivity $R = R(\omega)$ of a
TlCl crystal driven by the TA wave of frequency $\nu_{\rm ac} =
25$\,MHz and 125\,MHz is plotted in Figs.\,4\,(a) and (b),
respectively. In this case, in Eqs.\,(\ref{ham}) and
(\ref{macroP}) we put $m_3 = 0$, and only even-order
TA-phonon-assisted transitions occur. The Bragg signals $n=-2$,
due to $|r_{-2}|^2 \propto I_{\rm ac}^2$, are indicated in
Fig.\,4\,(a) by arrows, for the transitions within the LP and UP
branches, respectively. Similarly to the previous case (CuCl), for
THz light frequency $\omega$ close to $\Omega_{\rm TO}$ the
AW-induced change of $R$, $\Delta R = R - R^{(0)} \sim I_{\rm
ac}^2$ [see Fig.\,4\,(b)], is mainly due to the $n=0$ Bragg
replica. The quartic nonlinearity leads to the Stark blue shift by
$2 m_4 I_{\rm ac}$ of the TO-phonon frequency, according to
Eqs.\,(\ref{ham}) and (\ref{macroP}), as is clearly seen in
Figs.\,4\,(a) and (b). The Stark shift $\sim 0.1-0.2$\,THz has a
rather sharp contrast on the blue side of the THz reflectivity
[see inset in Fig.\,4\,(b)].

\begin{figure}[t]
\includegraphics[angle=270,width=0.5\linewidth]{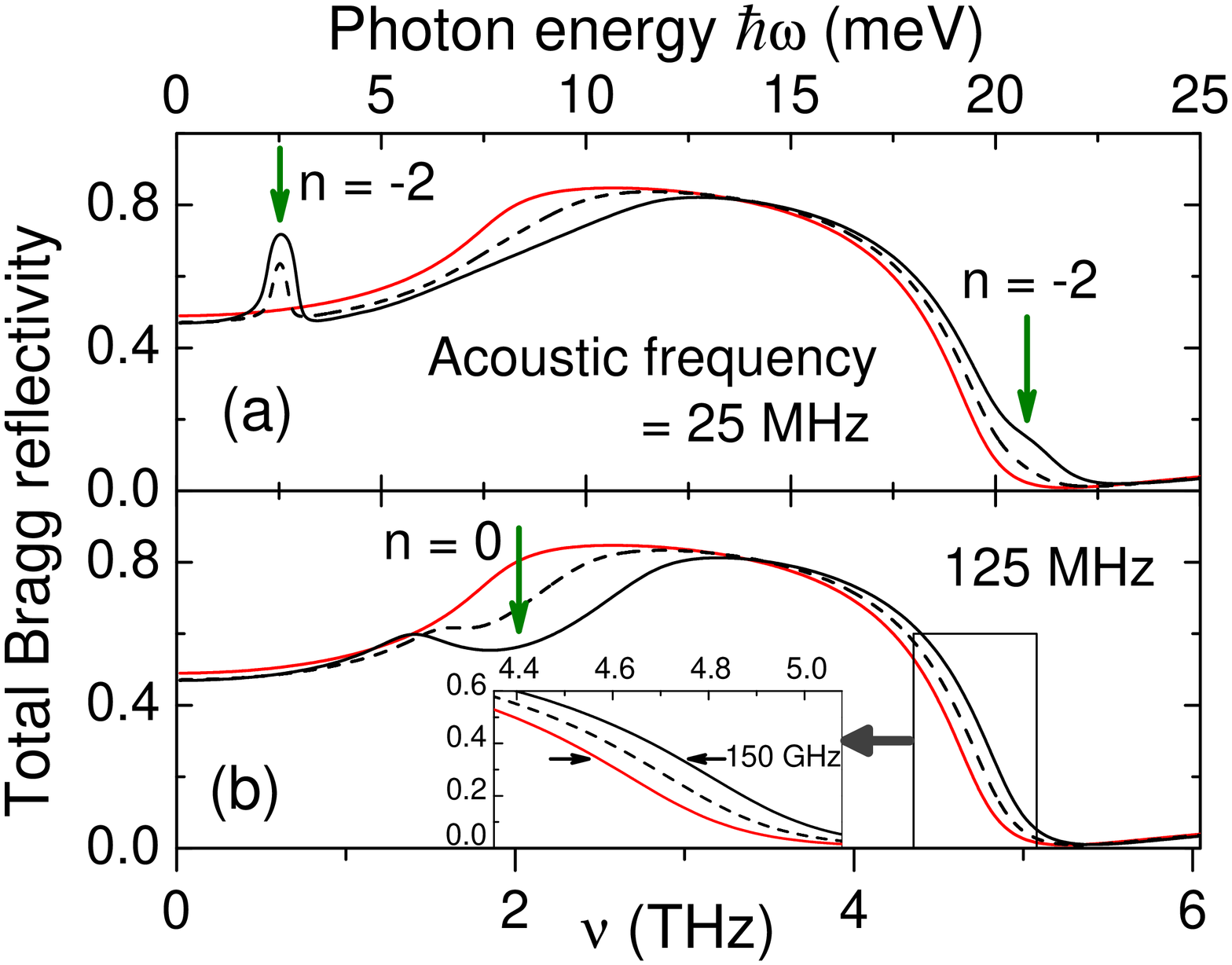}
\caption{(Color online) The total Bragg reflectivity $R =
R(\omega)$ of a TlCl crystal driven by the TA wave of frequency
$\nu_{\rm ac} = 25$\,MHz (a) and 125\,MHz (b), $I_{\rm ac} =
100$\,kW/cm$^2$ (black dotted lines) and 200\,kW/cm$^2$ (black
solid lines). The acoustically-unperturbed THz spectrum $R^{(0)} =
R^{(0)}(\omega)$ is shown by the red lines. Insert: The
acoustically-induced Stark shift of the TO-phonon line. $\hbar
\Omega_{\rm TO} = 7.81$\,meV, $\hbar \Omega_{\rm R} = 17.97$\,meV,
and $\hbar \gamma_{\rm TO} = 0.92$\,meV.}
\end{figure}

The acoustically-induced modulation of infrared polaritons has to
be particularly strong for ferroelectric soft TO-phonons (e.g., in
LiTaO$_3$ and LiNbO$_3$~\cite{Bakker1998} and bismuth titanate
\cite{Kojima2003}). In this case, a multi-well local potential for
the displacive ferroelectric mode has a considerable
low-wavevector component and therefore yields large values of
$V_3$ and $V_4$. Far-infrared optical phonons ($2-10$\,meV) in
zirconium tungstate (ZrW$_2$O$_8$) indicate anomalously high
anharmonicity~\cite{Hancock2004,Chaplot2005}. The normal modes
associated with soft TO-phonons in this negative thermal expansion
compound are a mixture of librational and translational motion.
The latter strongly couples with acoustic phonons giving rise, as
we foresee, to manipulation of the THz polaritons by using
utrasound waves of modest $I_{\rm ac}$.

We thank S.~G. Tikhodeev and R. Zimmermann for valuable
discussions. This work was supported by RS (Grant JP0766306),
EPSRC and WIMCS.

\end{document}